\documentclass[preprint,showpacs,preprintnumbers,amsmath,amssymb]{revtex4}
\usepackage{amsmath}
\usepackage{amsfonts}
\usepackage{hyperref}
\usepackage{tabularx}
\usepackage{graphicx}
\usepackage{bm}
\DeclareGraphicsExtensions {jpg,pdf}
\begin{document}

\title{Broadband impedance of the \emph {NESTOR} storage ring}
\thanks{Work partially supported by NATO grant SfP 977982}

\author{V.P. Androsov, P.I. Gladkikh, A.M. Gvozd, I.M. Karnaukhov, Yu.N. Telegin}
\email{telegin@kipt.kharkov.ua}
\affiliation{National Science Center\\
Kharkov Institute of Physics and Technology, Kharkov, Ukraine}%

\date{\today}

\headheight=14pt

\begin{abstract}
We have estimated contributions from the lossy and inductive vacuum chamber components
to the broadband impedance of the NESTOR storage ring by using analytical formulas. As was
expected considering the small ring circumference (15.44m), the main contributions both to the longitudinal impedance $Z_{\parallel}/n$  and the loss factor $k_{loss}$ come from the RF-cavity.

Cavity impedance was also estimated with CST Microwave Studio (CST Studio Suite$^{TM}$ 2006) by simulating coaxial wire method commonly used for impedance measurements. Both estimates agree well. The upper limit of impedance of elliptic holes in the vacuum chamber of dipole magnet were also obtained with this approach.

We have also evaluated the bunch length in NESTOR taking the conservative estimate of 3 Ohm
for the ring broadband impedance and have found that the bunch length $\sigma_z$= 0.5 cm could be obtained for the designed bunch current of 10 mA and RF-voltage of 250 kV.
\end{abstract}

\pacs{29.20.Dh, 29.27.Bd}
\maketitle

\section*{ Introduction}

At the design stage of any storage ring it is customary to study the effects of beam
interaction with vacuum chamber because these effects determine, for the
most part, the bunch parameters and limit the stored beam current. Since pioneer work by Sessler
and Vaccaro \cite{Sessler67} the effects of beam interaction with  a vacuum chamber (or a beam-pipe)
are treated in the frequency domain in terms of coupling impedances: longitudinal impedance,
$Z_{\parallel}(\omega)$ and transverse impedance $Z_{\perp}(\omega)$. Generally, the beam current
is characterized with higher moments which describe the dynamics of charge distribution
in the bunch, and one has to match to all moments m=1,2,3... the corresponding impedances
$Z_{\parallel}^m(\omega)$ and $Z_{\perp}^m(\omega)$ (see, for example, \cite{Chao93}). In practice
only impedances with m=0,1 which describe rigid bunch movement are studied
in order to optimize the design of beam-pipe components and to consider the forthcoming beam instabilities and their possible cures. In this paper we consider the longitudinal broadband
impedance which determines the bunch length in the NESTOR ring.

The longitudinal impedance is related to longitudinal beam component (m=0), which for a point charge
is $I_z=I_0\delta(x-x_1)\delta(y-y_1)\exp(-i\frac{\omega}{c}z)$, and can be expressed in
the following way:
\begin{equation} \label{c1}
Z_{\parallel}(\omega)\equiv Z_{\parallel}^{m=0}(\omega)=-\dfrac{1}{I_0}
\int\limits_{-\infty}^{\infty}E_z e^{ -j \omega c/z}dz \;,
\end{equation}
where $E_z $ is the longitudinal component of the electric field along the beam axis, $c$ is
velocity of light. Minus in  Eq.\eqref{c1} imply that  the induced field is retarding
(shifted in phase by $180^0$).

For convenience, one usually consider broadband (BB) ring impedance, that corresponds to all low-Q
components of the beam chamber, and narrow-band resonances originated from high-Q
elements, first of all, RF-cavities. RF-cavities along with fundamental (accelerating) mode
$TM_{010}$ have a whole spectrum of higher order modes (HOM's) which can drive coupled-bunch
instabilities (CBI) \cite{Laclare87}. Cavity HOM's as a possible origin of CBI in the
NESTOR were discussed in the previous paper \cite{telegin07}.

Longitudinal broadband impedance is usually normalized to a harmonic number $n=\omega/\omega_0$,
where $\omega_0$ is a rotation frequency, and is expressed as
$Z_{\parallel}/n$. For many beam-pipe elements $Z_{\parallel}(\omega)$ is inductive,
so it is supposed that $Z_{\parallel}(\omega)/n$ is constant in a wide frequency range.
$Z_{\parallel}/n$ is used for estimation of single-bunch instability thresholds with help
of half-empirical criteria obtained in low-frequency approximation ($\omega\ll c/\sigma_z$,
where $\sigma_z$ is a bunch length). Obtained in this approach, Keil-Schnell-Boussard criterion
\cite{Boussard75} relates $Z_{\parallel}/n$ to the microwave instability threshold:
\begin{equation} \label{c3}
\left|\dfrac{Z_{\parallel}}{n}\right|<\dfrac{2\pi \alpha (E_0/c)\delta_E^2}{I_{peak}}\;,
\end{equation}
where $\alpha$ is the momentum compaction factor, $E_0/e$ is the electron beam energy and $\delta_E$
is the relative beam energy spread. $I_{peak}$ represents the bunch peak current which for the gaussian bunch of length $\sigma_z$ is $I_{peak}=I_{av}\surd{2\pi}R/\sigma_z$, where $I_{av}$ is the average bunch
current and $R$ is the average radius of the ring. This approximation is good for a medium bunch
length: $\sigma_z\leq b$, where $b$ is the characteristic transverse size of the beam pipe.
For short bunches ($\sigma_z\ll c/\omega_r$) one has to use in  Eq.\eqref{c3} instead of the low-frequency impedance
$Z_{\parallel}/n$ the effective impedance ${(Z_{\parallel}/n)_{eff}}$, which is averaged
over bunch frequency spectrum, thus taking into account that at high frequencies the
bunch does not interacts with the low-frequency part of broadband impedance. For short gaussian bunches  a simple relation can be used:
$(Z_{\parallel}/n)_{eff}\approx2(\omega_r\sigma_z/c)^2\cdot\left|Z_{\parallel}/n\right|$

The real part of BB impedance determines beam energy losses. Parasitic energy loss per turn due to bunch-environment interaction is proportional to the bunch charge squared ($\Delta E=-k_{loss}q^2$), where factor $k_{loss}$ is a loss parameter which for gaussian bunch can be presented as follows:
\begin{equation} \label{c6}
k_{loss}=\dfrac{1}{2\pi}\int\limits_{-\infty}^{\infty} ReZ_{\parallel}(\omega)
\exp(-\dfrac{\omega^2\sigma_z^2}{c^2})dz\;,
\end{equation}

Below the cut-off frequency of a beam-pipe,  $\omega_{cut-off}$, broadband impedance is an additive value and the sum of contributions from various vacuum chamber components (so called "impedance budget") is usually evaluated in order to minimize this value. For the frequency-dependent components
(like resistive wall) impedance is evaluated at the roll-off frequency, $\omega_{roll-off}$
(the frequency at which the bunch spectrum power density $|\tilde{\rho}(\omega)|^2$ is one half of
its peak value).  For gaussian bunch $\omega_{roll-off}=c\sqrt2/\sigma_z $.

Total parasitic energy losses have to be considered in designing of heat-removing circuits and in
developing of the RF-system as a whole. The latter is especially important for low-energy rings like
NESTOR, in which the parasitic losses exceed synchrotron radiation losses and define the
synchronous phase of the beam.

In this paper we have evaluated contributions from various beam-pipe components to
the ring longitudinal BB impedance. Because of unavailability of 3D time-domain codes
like GdfidL\cite{GdfidL08} or new version of CST Studio Suite$^{TM}$ \cite{CST09}, which can directly calculate $Z_{\parallel}/n$, we had to restrict ourselves to analytical approaches. In most cases the latter give satisfactory results. We have also used the available version CST Studio Suite$^{TM}$ 2006 to evaluate the impedance of the elliptical holes in vacuum chambers of dipole magnets, which couldn't be estimated analytically.\\
\section*{1. The BB impedance of the \emph {NESTOR}. Analytical approach.}
 The next ring components can give a substantial contribution to the ring broadband impedance:

  -- resistive wall,

  -- RF-cavity;

  -- pumping slots in dipole vacuum chambers;

  -- beam pick-up's;

  -- holes in the laser-electron beam crossing chamber;

  -- RF-liners (bellows)

  -- welding joints.
\subsection*{1.1. The resistive wall impedance}

The resistive wall coupling impedance of a beam-pipe of elliptical cross section with major and minor
semi-axes $a$ and $b$  can be evaluated with the following equation \cite{Gluckstern93}:

\begin{equation} \label{c7}
\dfrac{Z_{\parallel}^{RW}}{n}=\dfrac{Z_0\delta}{2b}\dfrac{L}{2 \pi R}(1+j)G_0(q)\;,
\end{equation}
where $L$ is the pipe length, $\delta=(2c\rho/\omega Z_0)^{1/2}$  is the skin depth of the wall material
whose specific resistance is $\rho$, $Z_0$ is the impedance of free space.
$G_0(q)$ is a function of the parameter $q=(a-b)/(a+b)$ and is calculated via elliptic integrals
and Jacobi elliptic functions. For NESTOR $q$=0.49, function $G_0(q)\approx1$ and the resistive wall
impedance is given by a simple equation: $Z_{\parallel}^{RW}=1.35(1+i)\sqrt{n}$  [Ohm]. For the bunch
length $\sigma_z$=1 cm ($\omega_{roll-off}\approx$ 25GHz) the resistive wall impedance
$Z_{\parallel}^{RW}/n$ amounts approximately to 0.13 Ohm.

The estimate  of $k_{loss}^{RW}$ according to  Eq.\eqref{c6} gives 0.06 V/pC for bunch length
$\sigma_z$=1 cm.
\subsection*{1.2. The RF-cavity}

The contribution from the RF-cavity to the ring broadband impedance can be estimated by using the following equation \cite{Chae03}:

\begin{equation} \label{c8}
\dfrac{Z_{\parallel}^{HOM}}{n}=\sum_i (\dfrac{R}{Q})_i \cdot(\dfrac{\omega_0}{\omega_i})\;,
\end{equation}
where $\omega_i$ and $(R/Q)_i$ are the resonance frequency and R/Q factor for the cavity mode with mode index $i$. The sum is taken over all HOM's with $\omega_i<\omega_{cut-off}$.
HOM's parameters were calculated with ANSYS code  \cite{ANSYS04} using
approach described elsewhere \cite{telegin07}. The estimation gives $Z_{\parallel}^{HOM}/n$=1.40 Ohm.

Cavity HOM's give a significant contribution to the total loss factor. This contribution was estimated as sum of energy losses over all cavity HOM's:
\begin{equation} \label{c9}
k_{loss}^{HOM}=\sum_i (\dfrac{R}{Q})_i\cdot \omega_i\cdot\exp{-\dfrac{\omega^2\sigma_z^2}{c^2}} \;,
\end{equation}

It gives $k_{loss}^{HOM}$=0.48 V/pC for bunch length $\sigma_z$=1 cm and 0.54 V/pC for
$\sigma_z$=0.5 cm. Energy loss at fundamental mode doesn't depend on bunch length in this range
of $\sigma_z$ and amounts to 0.54 V/pC.

\subsection*{1.3. Beam position monitors}

\begin{figure}
\includegraphics [width=\columnwidth] {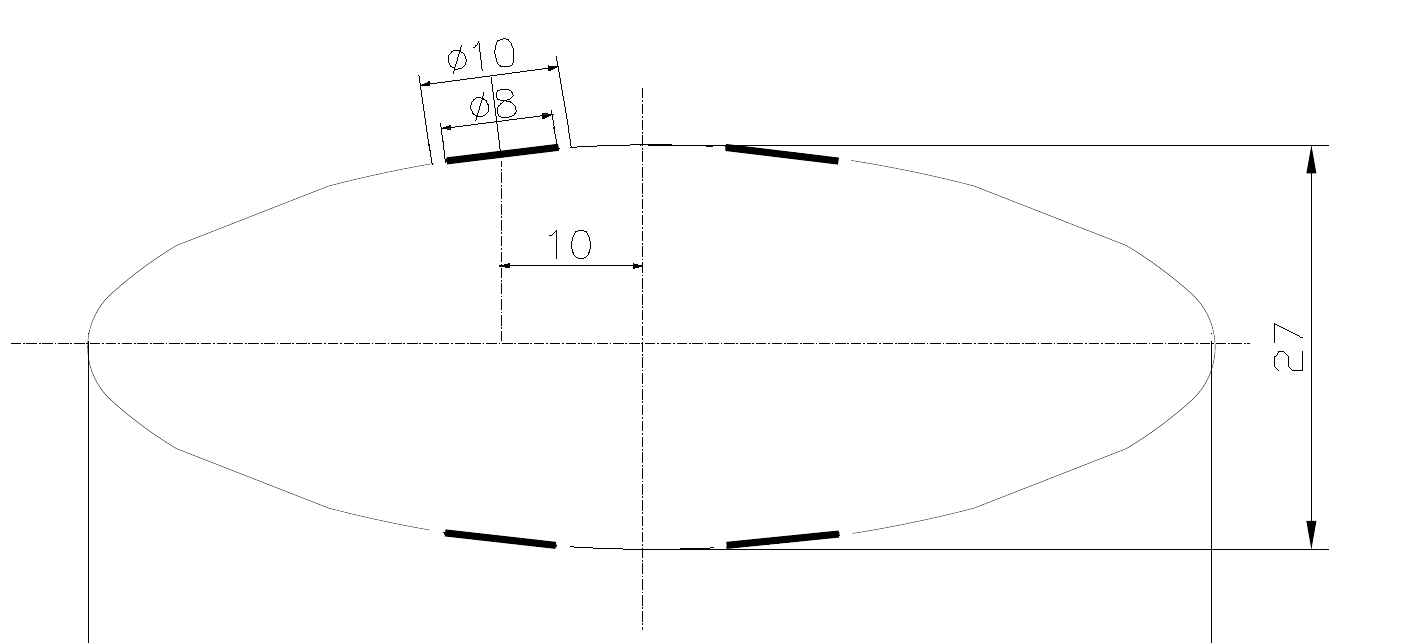}
\caption {\label{fig1} Beam pipe cross section at pick-up location }
\end{figure}

The beam position monitor (pick-up) represents four electrostatic electrodes (buttons)
placed on the inside surface of the beam-pipe and separated from the latter by narrow annular slots.
The beam-pipe cross section at pick-up location is given in Fig.~\ref{fig1}.
Impedance of one button can be estimated as impedance of an annular slot in the elliptic
beam-pipe. The longitudinal impedance of of a hole in an elliptic beam-pipe in static approximation
(hole dimensions are small compared with the
wavelength) is given by \cite{Gluck92}:

\begin{subequations}\label{c10}
\begin{align}
&Z_{\parallel}^{hole}=\dfrac{j\omega Z_0}{8\pi^2c}\cdot \dfrac{1}{a^2+b^2}\cdot
\dfrac{(\psi_v-\chi)\cdot Q_0^2(v)}{\sinh^2{u_0}+\sin^2{v}}\\
&Q_0(v)=\dfrac{2K(k)}{\pi}\cdot \dfrac{k'}{dn(\overline{v},q)}\;,
\end{align}
\end{subequations}
where $\psi_v$ and $\chi$ are the susceptibility and polarizability of the hole; $v$ is an azimuthal  elliptic coordinate; $u_0$ is related to ellipse semi-axes via: $\cosh{u_0}=a/\sqrt{a^2-b^2}$;
$dn(\overline{v},q)$ is Jacobi elliptic function of the argument $\overline{v}=2K(k)v/\pi)$; $K(k)$
is the complete elliptic integral of the first kind; $k'=\sqrt{1-k^2}$ and $q$ is defined above.
The azimuthal elliptic coordinate of the button center is $v=75.3^0$ and $Q_0(v)$=2.58.
\vspace{3mm}

The susceptibility $\psi$ and polarizability $\chi$ of a narrow annular slot
($w=r_{ext}-r_{int}\ll r_{ext}$, where $r_{ext}$ and $r_{int}$ are external and internal
radii of the slot) in a thin wall are given by \cite{Kurennoy95a}:
\begin{subequations}\label{c11}
\begin{align}
&\psi=\dfrac{\pi^2r^2_{ext}r_{int}}{\ln(32r_{ext}/w)-2}\\
&\chi=-\dfrac{1}{8}\pi^2w^2(r_{ext}+r_{int})\;,
\end{align}
\end{subequations}

Calculations give $Z_{\parallel}^{BPM}/n=1.1\cdot10^{-2}$ Ohm for one pick-up (4 buttons).
It should be noted that the value of pick-up impedance obtained as a difference of impedances of two
circular holes with radii $r_{ext}$ and $r_{int}$ \cite{Heifets96} is four times less than
the value  given above.

For estimation of the real part of the pick-up impedance we used the formula obtained
for a circular beam-pipe \cite{Kurennoy95} and modified in accordance with Eq.\eqref{c10}:
\begin{equation}\label{c12}
ReZ_{\parallel}^{hole}\!\!=Z_0\!\left(\dfrac{\omega}{c}\right)^4\!\!
\dfrac{(\psi_v^2+\chi^2)\cdot Q_0^2(v)}{96\pi^3\left[b^2+(a^2-b^2)\sin^2{v}\right]},
\end{equation}
where $Q_0(v)$ is defined by Eq.\eqref{c10}. The loss parameter is given by:
\begin{equation}\label{c13}
k_{loss}^{hole}=\dfrac{Z_0\sqrt{\pi}}{16\pi^4\sigma_z^5}\cdot
\dfrac{(\psi_v^2+\chi^2)\cdot Q_0^2(v)}{b^2+(a^2-b^2)\sin^2{v}}\;
\end{equation}

Calculation gives $k_{loss}^{BPM}=6.8\cdot 10^{-5}\cdot\sigma_z^{-5}[cm]$ V/pC for one pick-up.

\subsection*{1.4. Welding joints}

The adjacent sections of the stainless-still beam-pipe in NESTOR are joined together by welding
for which operation there are membranes at section ends. After welding between two joined
sections an annular gap $\sim$ 1.5mm wide and $\sim$ 3mm deep is formed. The contribution
of this gap to the ring broadband impedance has been estimated with the expression \cite{Bane88}:
\begin{equation}\label{c14}
\dfrac{Z_{\parallel}^{gap}}{n}=\dfrac{\omega_0Z_0g\Delta}{2\pi c a_{eff}}\;,
\end{equation}
where $g$ and $\Delta$ are the gap length and depth, correspondingly, $a_{eff}=\sqrt{ab}$ is the effective radius.
We estimated $Z_{\parallel}^{gap}/n$ to be $\sim 4.8\cdot 10^{-3}$ Ohm per joint.

\subsection*{1.5 Bellows liner }

To match the environment seen by the beam in bellows with circular cross section to that of the elliptic beam pipe the RF-liners will be used. The impedance of such a liner can be estimated as impedance of many identical slots which are equally spaced around the azimuthal perimeter \cite{Gluck92}:
\begin{subequations}\label{c15}
\begin{align}
&\dfrac{Z_{\parallel}^{liner}}{n}=\dfrac{jZ_0\eta (\psi_v-\chi)}{(a+b)}\cdot G_0(u_0)\\
&G_0(u_0)=\dfrac{\exp (u_0)}{4\pi}\cdot \int\limits_{0}^{2\pi}\dfrac{Q_0^2(v)dv}{[sinh^2u_0+sin^2v]^{1/2}}\;,
\end{align}
\end{subequations}
where $\eta=P/2\pi RC$, $P$ is the number of slots, $C$ is the circumference of the beam pipe cross section. For 22 slots, 2mm wide and 80mm long, the calculations give $Z_{\parallel}^{liner}/n= 2.1\cdot10^{-4}$  Ohm for one liner.

\section*{2. The broadband impedance of the \emph {NESTOR} ring. Simulations.}

For some beam pipe elements the BB impedance couldn't be estimated analytically. Such are pumping slots in the vacuum chamber of dipole magnet, which have transverse dimensions comparable to those of the beam pipe. We have tried to estimate their impedance by simulating coaxial wire method, widely used for impedance bench measurements  \cite{Caspers98}, with available code pack CST Studio Suite$^{TM}$ 2006. This version includes transient solver that can be used for S-parameter calculation.

The basic concept of this method relies on substituting the beam by a thin wire and thus simulating the fields of ultrarelativistic beam on the beam pipe wall by the propagation of TEM mode in the transmission line so formed. In the transmission line framework a single, lumped wall impedance $Z_W$ can be expressed in terms of S-matrix coefficients in a following way:\\
\begin{equation}\label{c16}
S_{11}=\frac{Z_W}{2R_c+Z_W} \quad \textrm{and}\quad S_{21}=\frac{2R_c}{2R_c+Z_W}   \;,
\end{equation}
where $R_c$ is characteristic impedance of the transmission line. In principle either coefficient can be used for impedance determination, but the transmission coefficient is applicable in more general configurations and is usually preferred in bench measurements.

The question to what degree the simulated impedance  $R_W$ approximate the BB impedance is raised from time to time \cite{Vaccaro99, Hahn04}. A good agreement between calculations and measurements has been shown for a simple cavity-like structure with a thin wire ($d$=0.75 mm) \cite{Vaccaro99}.

So as the first step we applied this approach to estimate the broadband impedance of the RF-cavity and to compare it to that obtained with analytical approach.

\subsection*{2.1 The RF-cavity}
\begin{figure}[t]
\includegraphics [bb=0 0 280 140] {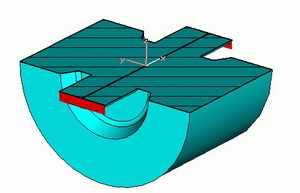}
\caption {\label{fig2} Cavity model used in calculations.}
\end{figure}

\begin{figure}[t]
\includegraphics [width=\textwidth] {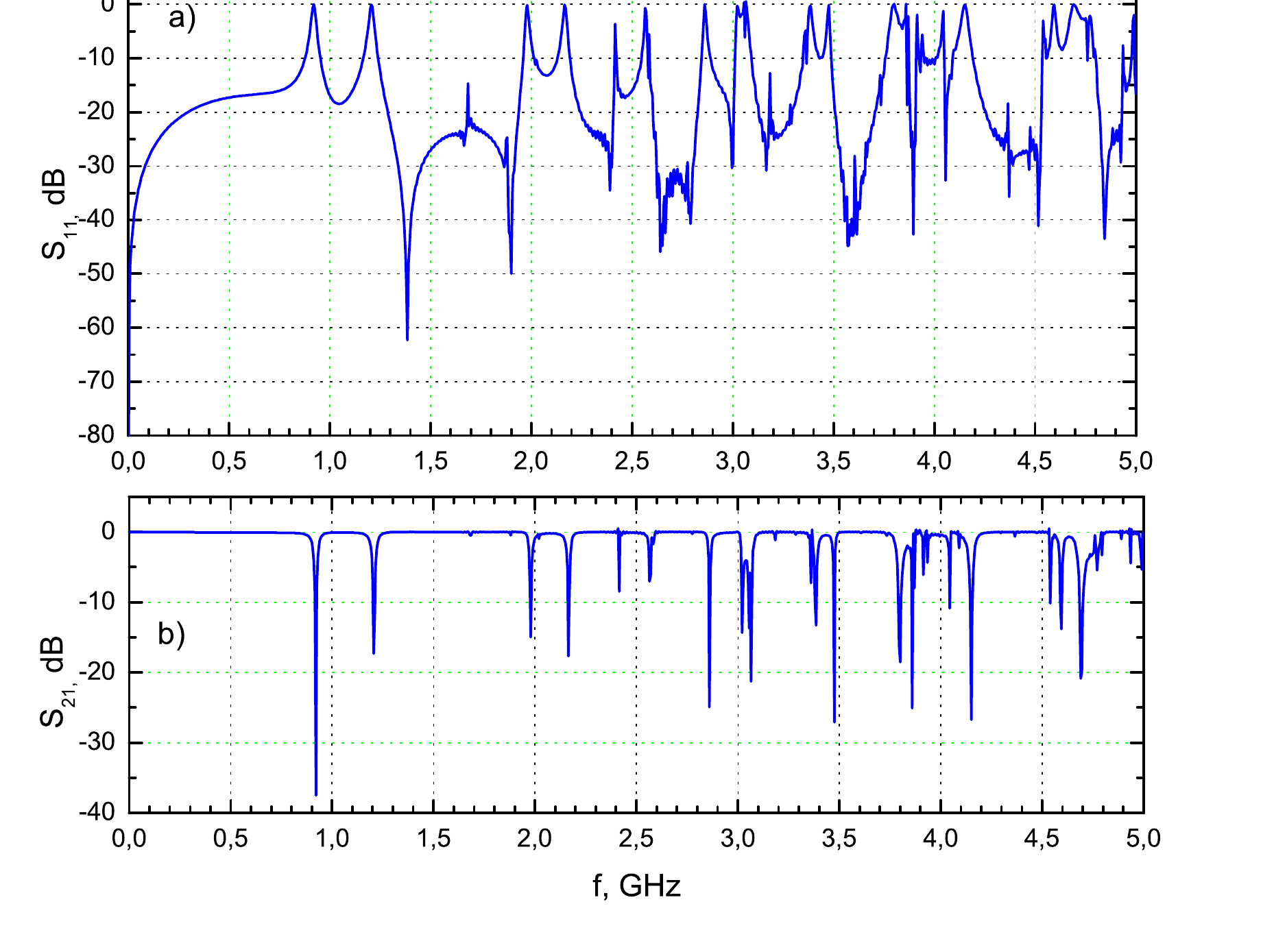}
\caption {\label{fig3} Scattering matrix parameters: a) $S_{11}$(reflection) and b) $S_{21}$(transmission)}
\end{figure}

The simplified  cavity model (without ports) used in calculations is shown in Fig.~\ref{fig2}.  The calculated S- parameters in the frequency range $1\div5$ GHz presented in Fig.~\ref{fig3}.
%

The analysis of the electromagnetic field distribution at resonance frequencies reveals:

1. The field patterns at resonance frequencies (minima in $S_{21}$) in peripheral region are similar to that of cavity modes. In particular, first resonance corresponds to $TM_{010}$ mode but the mode frequency is shifted to higher frequencies. The field in the output beam pipe is zero i.e. the cavity acts as a notch filter.

2. At frequencies corresponding to minima in $S_{11}$ the electromagnetic wave doesn't excite the cavity (field in peripheral region is zero) and the structure behaves as an usual coaxial transmission line.

Cavity impedance  obtained from $S_{21}$ with Eq.\eqref{c16} is presented in Fig. 4.
\begin{figure}[h]
\includegraphics [width=160mm] {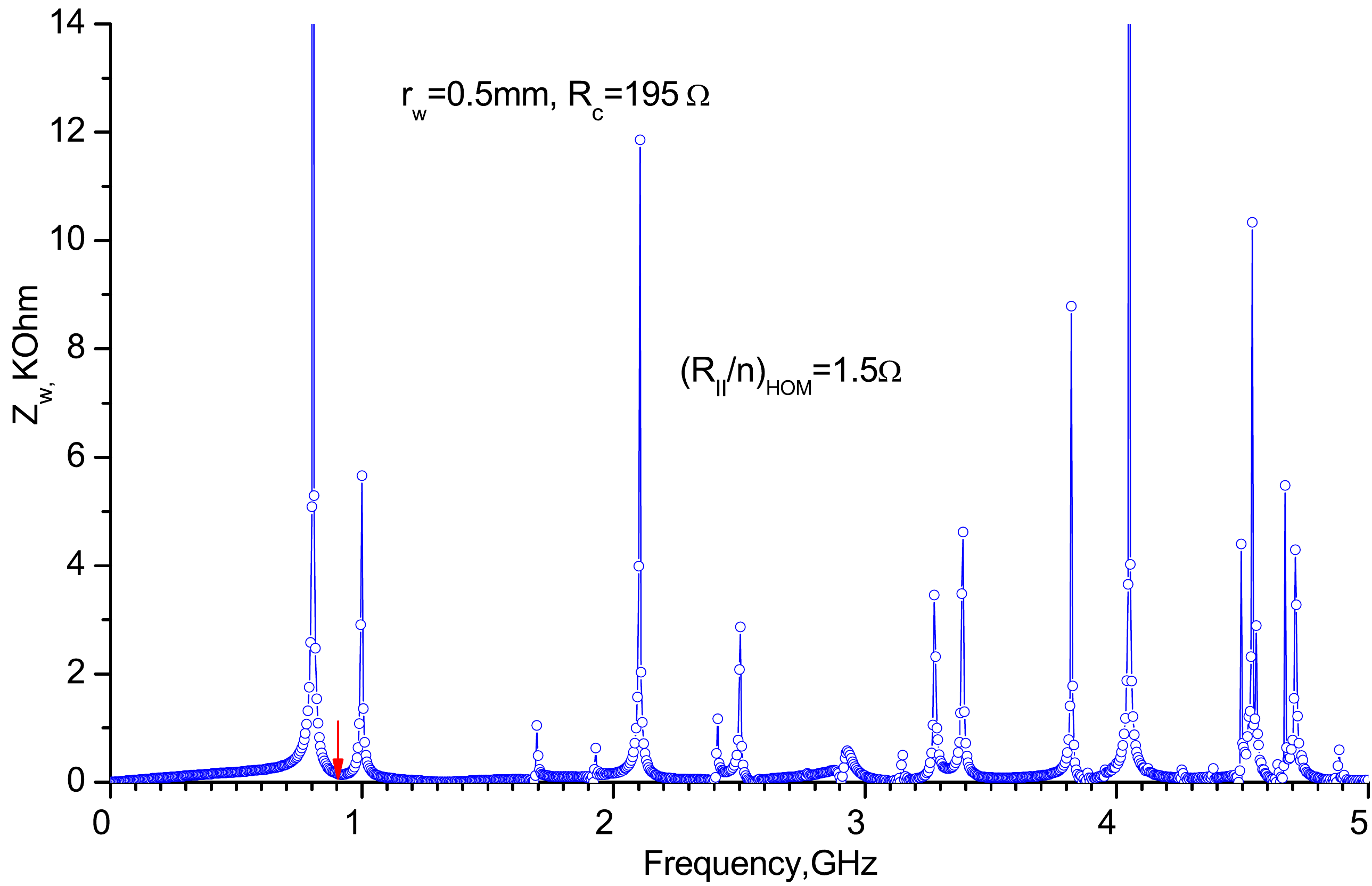}
\caption {\label{fig4} Cavity impedance.}
\end{figure}
%

Contribution to the ring broadband impedance from cavity HOM's was calculated with the following expression:\\
\begin{equation}\label{c17}
Z_{\parallel}^{HOM}/n=\frac{1}{\omega_{max}-\omega_c}\cdot\int \limits_{\omega_c}^{\omega_{max}}Z_W(\omega)\frac{\omega_0}{\omega}d\omega \;
\end{equation}
Bottom limit of integration $\omega_c=2\pi f_c$ was taken in the minimum between the first and the
second resonances (it is shown by an arrow in Fig.~\ref{fig4}).\\

The obtained value $Z_{\parallel}^{HOM}/n$=1.5 Ohm well agrees with the estimate obtained with analytical approach (see 3.1). It doesn't include contribution from propagating modes because $\omega_{cut-off}$= 5.9 GHz for the cavity elliptic beam pipes.
\subsection*{2.2 The dipole chamber}

\begin{figure}[h]
\includegraphics [bb=0 0 400 130] {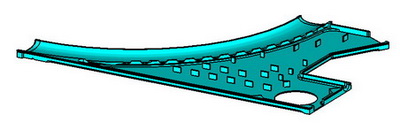}
\caption {\label{fig5} A vacuum chamber in the dipole magnet (bottom half) }
\end{figure}

\begin{figure}[h]
\includegraphics [bb=0 0 300 160] {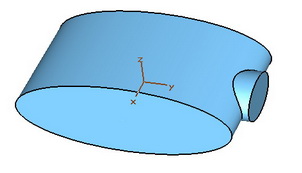}
\caption {\label{fig6} The model of the chamber fragment with a single hole used in simulations. }
\end{figure}

The cross section of the main vacuum chamber in the dipole magnet is similar to that presented in Fig. 1. The structure of the chamber is presented in Fig.~\ref{fig5}. The main chamber is pumped with an ion pump, mounted on antechamber, through 16 pumping slots of elliptic cross section (22x14 mm). Two chambers of total number of four have additional holes for electron beam injection or X-ray beam extraction. The simulated model, which presents the "straightened" segment of the chamber with a single hole, is presented in Fig.~\ref{fig6}.
\vspace{3mm}
%

To control the accuracy of obtained results we have made parallel simulations for similar elliptic segment without a hole. Impedance of the last is zero. Simulations have shown that obtained values of $S_{11}$ are less than $10^{-4}$ and are comparable to the level of numerical noise. The number of mesh points $N_c$ varied in the range 40000$\div$400000, at higher values of $N_c$ the simulations became unstable. Calculations were performed on the typical office PC and were CPU-bound, so we had no opportunity to improve simulated model.

\begin{figure}[h]
\includegraphics [width=\columnwidth] {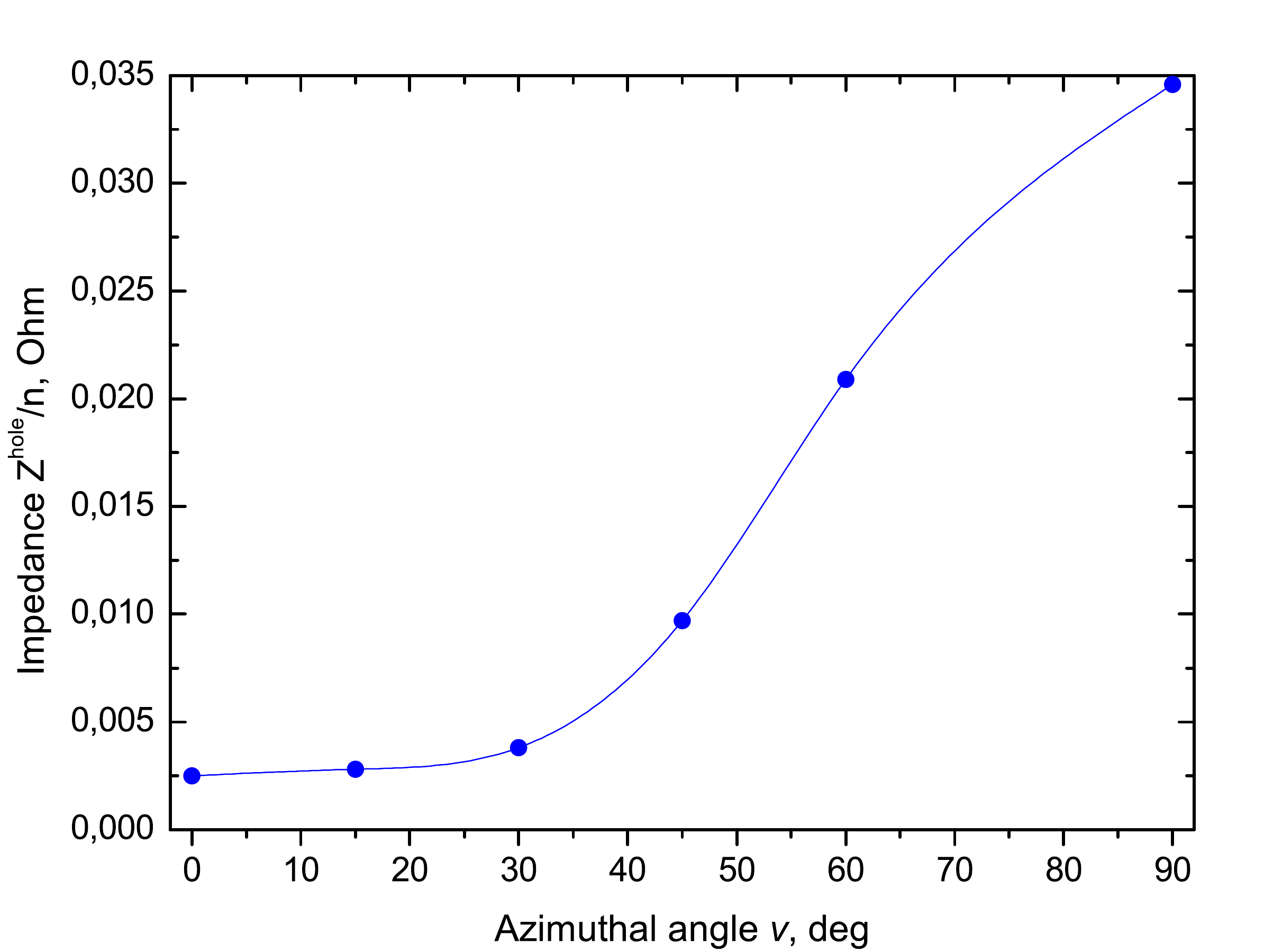}
\caption {\label{fig7} The hole impedance versus a hole azimuthal angle. }
\end{figure}

Nevertheless, we have evaluated the upper limit of hole impedance by using strong azimuthal dependance predicted by Eqs.\eqref{c10} for impedance of a small hole in the elliptic beam pipe with a substantial eccentricity. The elliptic hole (22x14 mm) in the elliptic pipe (79x27 mm) was placed at various azimuthal angles perpendicular to the tangent plane. The thickness of the wall in the hole center was constant. The broadband impedance was calculated with Eq.\eqref{c17}, averaging was done over the range of $0\div\omega_{max}$. The obtained dependence of $Z_{\parallel}^{hole}/n$ on elliptic azimuthal angle $v$ is presented in Fig.~\ref{fig7}. \\
%
 One can see that for azimuthal angles $v<30^0$ the calculated impedance is nearly constant. It can be explained by the fact that at low angles the hole impedance decreases to values, which are below the level of numerical noise. The figure also shows that the single hole impedance $Z_{\parallel}^{hole}/n>$0.003 Ohm can be quite reliably deduced from $S_{11}$-parameter in coaxial wire simulations. For reasons given above the value 0.05 Ohm can be considered as the upper limit of the impedance of the dipole chamber (16 holes). One should also keep in mind that such an additive estimate is valid only in low-frequency approximation.
%
%

%
%

%
\section*{3. The BB impedance of the \emph {NESTOR} ring. Results.}

The obtained results are summarized in Tables 1,2. As follows from the tables
for bunch length considered ($\sigma_z$=0.5, 1.0 cm) the main contributions both to longitudinal
broadband impedance and loss factor come from the RF-cavity. From a comparison of the loss factor
data for two bunch lengthes one can see how the contributions from other components are increased
with  bunch shortening while the contribution from the RF-cavity isn't changed essentially.

\renewcommand {\tabularxcolumn}[1]{m{#1}}
%
\begin{center}
\textbf{\emph{Table 1.}} {\it Longitudinal BB impedance budget}
\end{center}
\begin{tabular*} {\linewidth} {|l|c|c|c|}\hline
Component $\;\;\;\;\;\;\;\;\;$          &$\;\;\;\;\;\;\;$ Number  $\;\;\;\;\;\;\;$      &$\;\;\;\;\;\;\;$ $L$, nH (1 unit) $\;\;\;\;\;\;$  & $\;\;\;\;\;\;\;$ $|Z_{\parallel}/n|, \Omega$ (total) $\;\;\;\;\;\;$  \\
\hline
RF-cavity          &1               &                 &1.40     \\
Resistive wall     &1               &1.07             &0.13     \\
Dipole chamber     &4               &$<0.4$           &$<0.2$   \\
BPM                &4               &0.09             &0.044     \\
Welding joints     &8               &0.04             &0.04     \\
RF-liners          &4               &0.002            &0.0009     \\
\hline
Total                               &          &      &$<1.65$  \\
\hline
\end{tabular*}

%
\begin{center}
\textbf{\emph{Table 2.}} {\it Loss factors of beam pipe components}
\end{center}
\begin{tabular*} {\linewidth} {|l|c|c|c|}
\hline   &      &\multicolumn{2}{c|} {$k_{tot}$,V/pC } \\
\cline{3-4}
\raisebox{1.5ex}[0cm][0cm] {Component $\;\;\;\;\;\;\;\;\;\;\;\;\;\;$} &\raisebox{1.5ex}[0cm][0cm] {$\;\;\;\;\;\;\;$ Number $\;\;\;\;\;\;$} &$\;\;\;\;\;\;\;\;\;\;$ $\sigma_z$=1cm $\;\;\;\;\;\;\;\;\;\;\;$ &$\;\;\;\;\;\;\;\;\;\;$ $\sigma_z$=0.5cm $\;\;\;\;\;\;\;\;\;\;\;$ \\
\hline
RF-cavity      &1        &1.02               &1.08     \\
Resistive wall &1        &0.06               &0.17     \\
BPM            &4        &0.0003            &0.0088   \\
\hline
Total          &         &1.08               &1.26     \\
\hline
\end{tabular*}
\\

It should be mentioned that for the time being we have no estimates for two components:
i) the beam-pipe section for the crossing point of the electron and laser beams;
ii) the injection section (strip-line inflector).
The first element at the commissioning stage will be replaced with a
straight section. The injection section is essentially non-symmetric, it gives, presumably,
a substantial contribution to broadband impedance and so it requires to be studied with 3D
time-domain codes.

\section* {4. Turbulent bunch lengthening in the \emph {NESTOR} storage ring}

The turbulent bunch lengthening is observed at bunch currents higher some threshold value and is explained by coherent mode coupling arising through overlapping of mode frequencies at high bunch currents \cite{Sacherer77}. This effect displays itself as a longitudinal instability (microwave instability) and is accompanied by increase of the beam energy spread.

The threshold bunch current $I_{av}^{th}$ for a given value of $Z_{\parallel}/n$
can be obtained with Eq.\eqref{c3}. Above threshold the bunch length can be estimated with the
equation \cite{Hofmann80}:
\begin{equation}\label{c18}
\sigma_z^{tbl}=R\left(\sqrt{2\pi}\left|\dfrac{Z_{\parallel}}{n}\right|\dfrac{I_{av}}{hV_c
\sin{\Phi_s}}\right)^{1/3},
\end{equation}
where $h$ is the harmonic number, $V_c$ is the amplitude of RF-voltage and $\Phi_s$ is the synchronous
phase of the beam. One can see from Eq.\eqref{c18} that the bunch length doesn't
depend on the electron beam energy above the instability threshold.

We estimated  bunch lengths for a reasonable range of RF-voltages. NESTOR parameters used in these
calculations are presented in Table 3, the obtained results are given in Table 4.

In our calculations we have taken $Z_{\parallel}/n$=3 Ohm, that is almost two times the value given
in the table, in order to make compensation for the contributions from beam-pipe components not considered in the paper, first of all, the inflector. The values of the energy spread $\delta_E^{ibs}$
and bunch length $\sigma_z^{ibs}$ presented in Table 4 were calculated with DECA code
\cite{Zelinsky04}, which takes into account the effect of intrabeam scattering (IBS) -- electron-electron scattering in a bunch.
The threshold currents obtained with Eq.\eqref{c3} for these IBS-corrected bunch parameters are below
the goal value of 10 mA per bunch, so the  bunch will be lengthening in the ring. For the designed RF-voltage of 250kV estimation gives $\sigma_z^{tbl}$=0.5 cm.

\vspace{3 mm}
\begin{center}
\textbf{\emph{Table 3.}} {\it NESTOR ring parameters}
\end{center}
\begin{tabular*} {\linewidth} {|l|c|c|}
\hline
Parameter             &$\;\;\;\;\;\;\;\;\;\;$ Units $\;\;\;\;\;\;\;\;\;\;$            &$\;\;\;\;\;\;\;\;\;\;$value $\;\;\;\;\;\;\;\;\;\;$  \\
\hline
Average radius, $R$                           &m                  &2.46 \\
Harmonic number, $h$                          &                   &36   \\
Compaction factor, $\alpha$                    &                   &0.01 \\
Bunch current (maximal), $I_{av}$              &mA                 &10   \\
Ring broadband impedance, $Z_{\parallel}/n$ $\;\;\;\;\;\;\;\;\;\;\;\;\;\;\;\;\;\;\;\;\;\;\;$   &$\Omega$           &3.0  \\
Synchronous phase, $\Phi_s$                   &deg           &89   \\
\hline
\end{tabular*}

\begin{center}
\textbf{\emph{Table 4.}} {\it Bunch length in NESTOR}
\end{center}
\begin{tabularx}{165 mm}{|l|X|c|c|X|c|c|c|}
\hline &\multicolumn{3}{c|}{$E_0$=60 MeV}  &\multicolumn{3}{c|}{$E_0$=250 MeV}  & \\
\cline{2-7}
\raisebox{1.5ex}[0cm][0cm] {$V_c$, kV } &$\delta_E^{ibs}$  &$\sigma_z^{ibs}$, cm
&  $I_{av}^{th}$, mA  &$\delta_E^{ibs}$  &$\sigma_z^{ibs}$, cm &  $I_{av}^{th}$, mA
&\raisebox{1.5ex}[0cm][0cm] {$\sigma_z^{tbl}$, cm } \\
\hline
10   &$0.17\cdot10^{-2}$ & 1.32  &7.7   &$0.47\cdot10^{-3}$ & 0.71 &1.2   &1.46  \\
50   &$0.19\cdot10^{-2}$ & 0.67  &5.1   &$0.53\cdot10^{-3}$ & 0.36 &0.8   &0.85   \\
180  &$0.21\cdot10^{-2}$ & 0.40  &3.7   &$0.59\cdot10^{-3}$ & 0.21 &0.6   &0.56   \\
250  &$0.22\cdot10^{-2}$ & 0.34  &3.4   &$0.60\cdot10^{-3}$ & 0.18 &0.5   &0.50 \\
\hline
\end{tabularx}
\vspace{3 mm}

At the commissioning stage of the NESTOR project the RF-amplifier with output power of 1 kW is available.
It provides the accelerating voltage of 50 kV and the stored beam currents of up to 100 mA. We have no equipment to control a ring filling pattern, so all 36 RF-buckets will be filled up, and a maximal bunch current will be decreased down to 3 mA. For this case estimation gives $\sigma_z^{tbl}$=0.6 cm.
\section*{5.Conclusion}

The conservative estimate of the longitudinal broadband impedance of the NESTOR ring
($Z_{\parallel}/n$=3 Ohm) confirms the possibility of obtaining the bunch length of
0.5 cm with the goal bunch current of 10 mA for the designed RF-voltage of 250 kW.
At the commissioning stage we can obtain 3mA in a 0.6 cm bunch with the RF-voltage of 50 kV.

\bibliography{Impedances}

\end{document}